\pdfoutput=1
\documentclass[prb,twocolumn,showpacs,amsmath,amssymb,floatfix,superscriptaddress]{revtex4}

\usepackage{graphicx}
\usepackage{epsfig}
\usepackage{dcolumn}
\usepackage{bm}
\usepackage{float}

\begin{document}

\title{Spin-to-charge conversion in lateral and vertical topological-insulator/ferromagnet heterostructures with microwave-driven precessing magnetization}

\author{Farzad Mahfouzi}
\affiliation{Department of Physics and Astronomy, University of Delaware, Newark, DE 19716-2570, USA}
\author{Naoto Nagaosa}
\affiliation{RIKEN Center for Emergent Matter Science (CEMS), Wako, Saitama 351-0198, Japan}
\affiliation{Department of Applied Physics, University of Tokyo, Tokyo 113-8656, Japan}
\author{Branislav K. Nikoli\' c}
\email{bnikolic@udel.edu}
\affiliation{Department of Physics and Astronomy, University of Delaware, Newark, DE 19716-2570, USA}
\affiliation{RIKEN Center for Emergent Matter Science (CEMS), Wako, Saitama 351-0198, Japan}

\begin{abstract}
Using the charge-conserving Floquet-Green function approach to open quantum systems driven by external time periodic potential, we analyze how spin current pumped (in the absence of any dc bias voltage) by the precessing magnetization of a ferromagnetic (F) layer is injected {\em laterally} into the interface with strong spin-orbit coupling (SOC) and converted into charge current flowing in the same direction. In the case of metallic interface with the Rashba SOC used in  experiments  [Nature Comm. {\bf 4}, 2944 (2013)], both spin $I^{S_\alpha}$ and charge $I$ current flow within it where $I/I^{S_\alpha} \simeq$ 2--8\% (depending on the precession cone angle), while for F/topological-insulator (F/TI) interface employed in related experiments (arXiv:1312.7091) the conversion efficiency is greatly enhanced $I/I^{S_\alpha} \simeq$ 40--60\% due to perfect spin-momentum locking on the surface of TI. The spin-to-charge conversion occurs also when spin current is pumped {\em vertically} through the F/TI interface with smaller efficiency $I/I^{S_\alpha} \sim 0.001\%$, but with charge current signal being sensitive to whether the Dirac fermions at the interface are massive or massless.
\end{abstract}

\pacs{72.25.Mk,73.43.-f,73.40.-c}
\maketitle

One of the central goals of second generation spintronics~\cite{Awschalom2007} is to generate and manipulate
pure spin currents with no net charge flux. The pure spin currents make possible transport of information encoded in electron
spin with much less dissipation than generated using spin-polarized charge current of first generation
spintronics. However, their detection and measurement requires to convert them into conventional charge currents
and voltages. Over the past decade, the inverse spin Hall effect~\cite{Hirsch1999} (SHE)---where pure spin current
injected longitudinally into a material with spin-orbit coupling (SOC) induces~\cite{Hankiewicz2004a,Chen2012a} transverse charge current---has
emerged as the standard detector which has routinely been coupled to generators of pure spin currents like spin pumping by precessing
magnetization,~\cite{Saitoh2006,Mosendz2010,Ando2011a} nonlocal spin diffusion,~\cite{Valenzuela2006,Seki2008} direct SHE,~\cite{Brune2010,Abanin2011}
magnon-spin transmutation,~\cite{Kajiwara2010} and laser pulses.~\cite{Werake2011}

\begin{figure}[!]
\includegraphics[scale=0.30,angle=0]{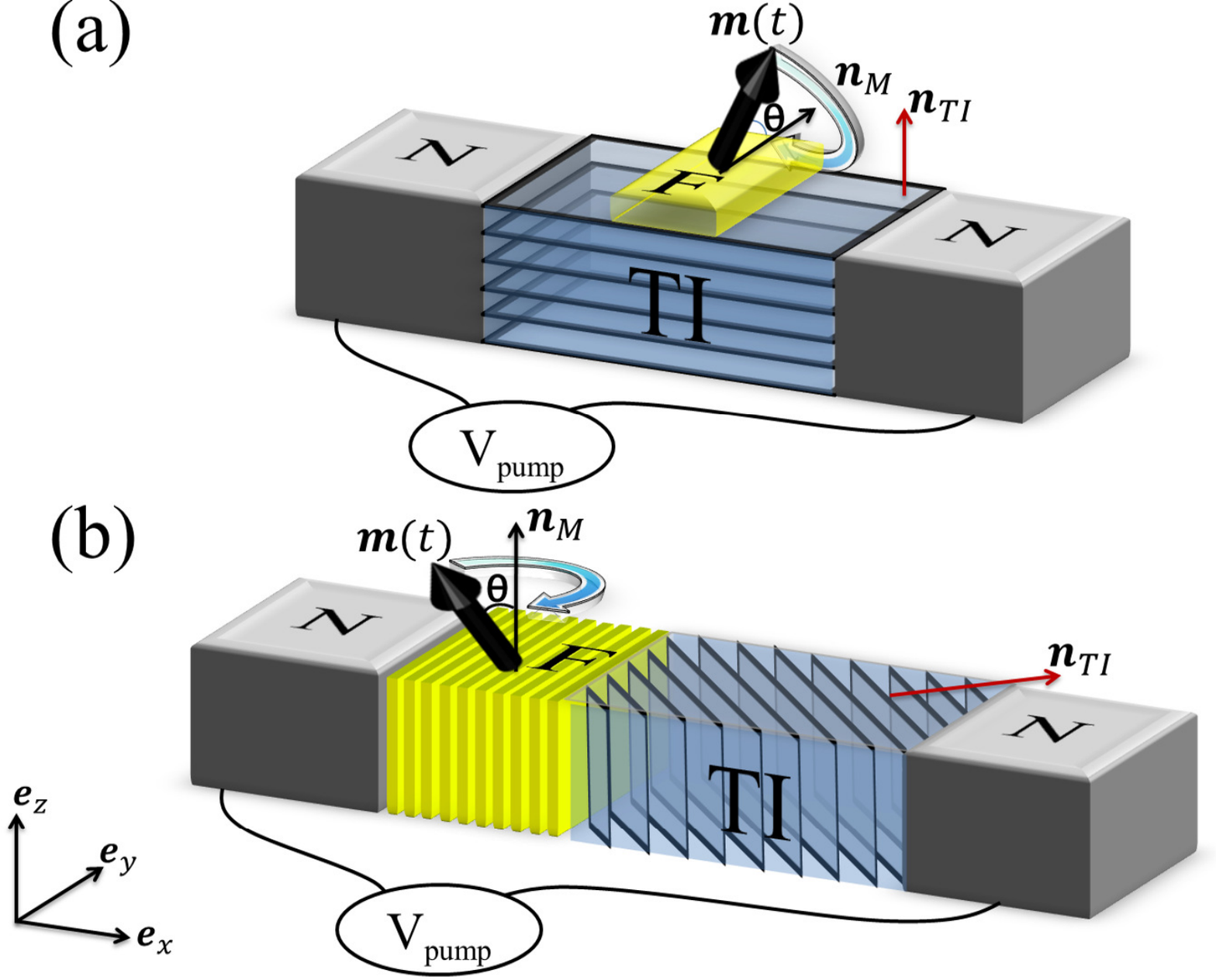}
\caption{(Color online) Schematic of (a) lateral and (b) vertical F/TI heterostructures whose precessing magnetization, driven by the absorption of microwaves of frequency $\omega$ under the FMR conditions, pumps pure spin current along the $z$- or $x$-axis, respectively, in the absence of any dc bias voltage between the semi-infinite N leads. The strong interfacial SOC converts pumped spins into charge current flowing along the $x$-axis in both setups, which is measured as the voltage signal $V_\mathrm{pump}$ in an open circuit. Besides lateral heterostructures of F/TI type in panel (a), we also study lateral F/2DEG heterostructures with conventional Rashba SOC at the interface for comparison. The unit vector $\mathbf{n}_M$ specifies the axis around which magnetization $M\mathbf{m}(t)$ is precessing, while the vector $\mathbf{n}_\mathrm{TI}$ is perpendicular to QLs comprising the 3D TI slab.}
\label{fig:fig1}
\end{figure}

The very recent experiments~\cite{Sanchez2013,Shiomi2013} on lateral heterostructures illustrated in Fig.~\ref{fig:fig1}(a)---where ferromagnetic (F) layer
with precessing magnetization driven by microwaves of frequency $\omega$ under the ferromagnetic resonance (FMR) condition is brought into a contact with a two-dimensional (interfacial) gas of either conventional electrons with parabolic energy-momentum dispersion (as formed at Ag/Bi interface~\cite{Sanchez2013}) described by the Rashba Hamiltonian~\cite{Winkler2003}
\begin{equation}\label{eq:rashba}
\hat{H}_\mathrm{2DEG} = \frac{\hat{\mathbf{p}}^2}{2m^*} + \frac{\alpha_\mathrm{R}}{\hbar}(\hat{\bm \sigma} \times \hat{\mathbf{p}}) \cdot \mathbf{e}_z,
\end{equation}
or massless Dirac electrons with linear energy-momentum dispersion on the surface of three-dimensional topological insulators (3D TIs)~\cite{Shiomi2013}
described by the Hamiltonian~\cite{Hasan2010}
\begin{equation}\label{eq:dirac}
\hat{H}_\mathrm{TI} = v_F (\hat{\bm \sigma} \times \hat{\mathbf{p}}) \cdot \mathbf{e}_z,
\end{equation}
both of which exhibit spin-momentum locking due to interfacial SOC term---have observed charge current $I$ along the $x$-direction
in the absence of any applied dc bias voltage. Here $\hat{\mathbf{p}}=(\hat{p}_x,\hat{p}_y)$ is the momentum operator, $m^*$ is the effective mass ($m^*/m_e \simeq 0.35$ at the Ag/Bi interface), $\hat{\bm \sigma} = (\hat{\sigma}_x,\hat{\sigma}_y,\hat{\sigma}_z)$ is the vector of the Pauli matrices and $v_F$ is the Fermi velocity. While Hamiltonians in Eqs.~\eqref{eq:rashba} and ~\eqref{eq:dirac} are quite similar, the main difference is that SOC term in Eq.~\eqref{eq:rashba} is a perturbation on the top of band kinetic energy, while in Eq.~\eqref{eq:dirac} it is the only term present.

\begin{figure}
\includegraphics[scale=0.34,angle=0]{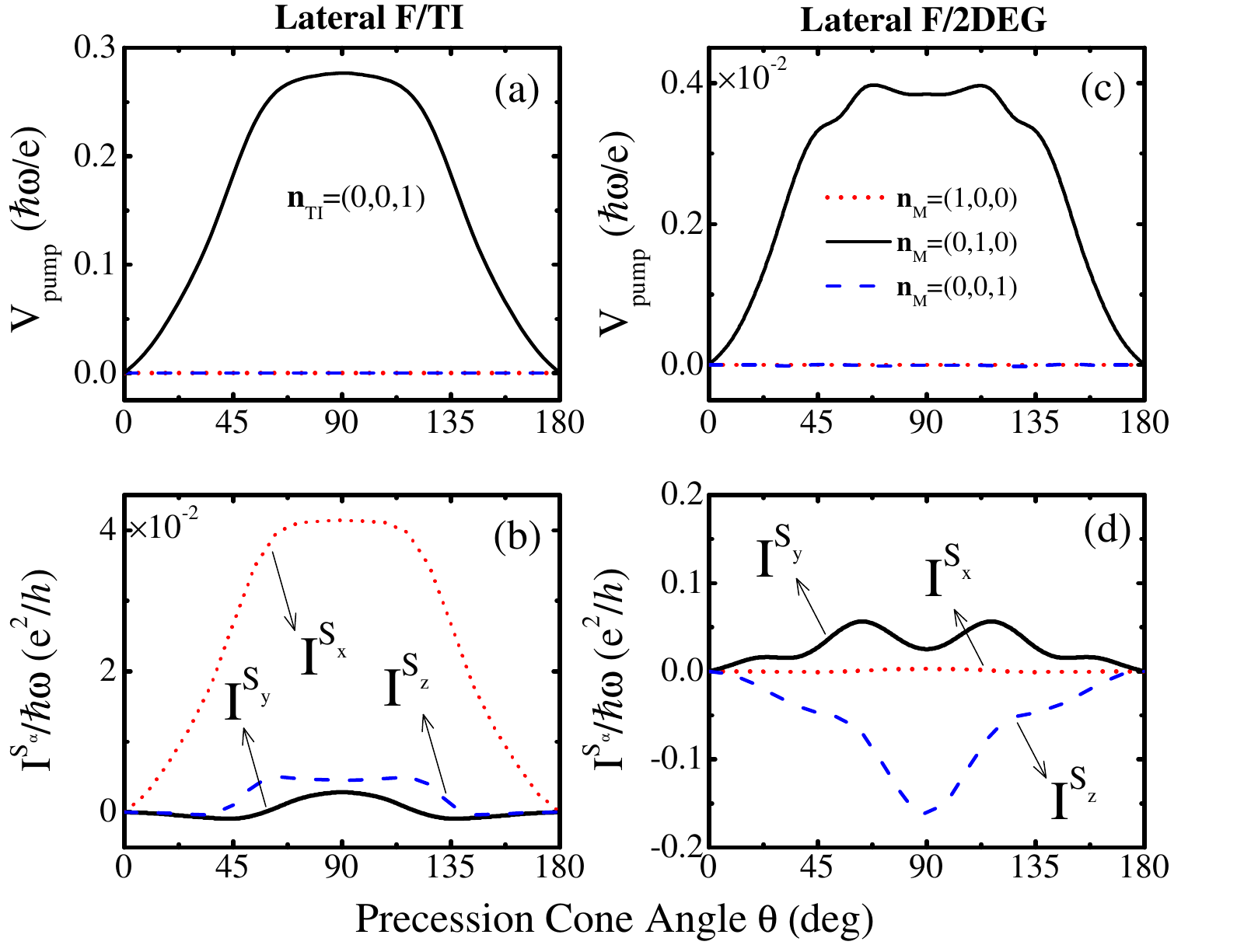}
\caption{(Color online) (a), (c) The angular dependence of the dc pumping voltage in lateral F/TI and F/2DEG heterostructures illustrated in Fig.~\ref{fig:fig1}(a)  for different orientation of the axis $\mathbf{n}_M$ around which magnetization is precessing with a cone angle $\theta$. Panels (b) and (d) show the spin current $I^{S_\alpha}$ which accompanies the charge current $I = V_\mathrm{pump} G$ in (a) and (c), respectively, where both $I^{S_\alpha}$ and $I$ flow along the $x$-axis in Fig.~\ref{fig:fig1}(a). Note that charge current is non-zero only when magnetization is precessing around the $y$-axis in Fig.~\ref{fig:fig1}(a).}
\label{fig:fig2}
\end{figure}
\begin{figure}
\includegraphics[scale=0.33,angle=0]{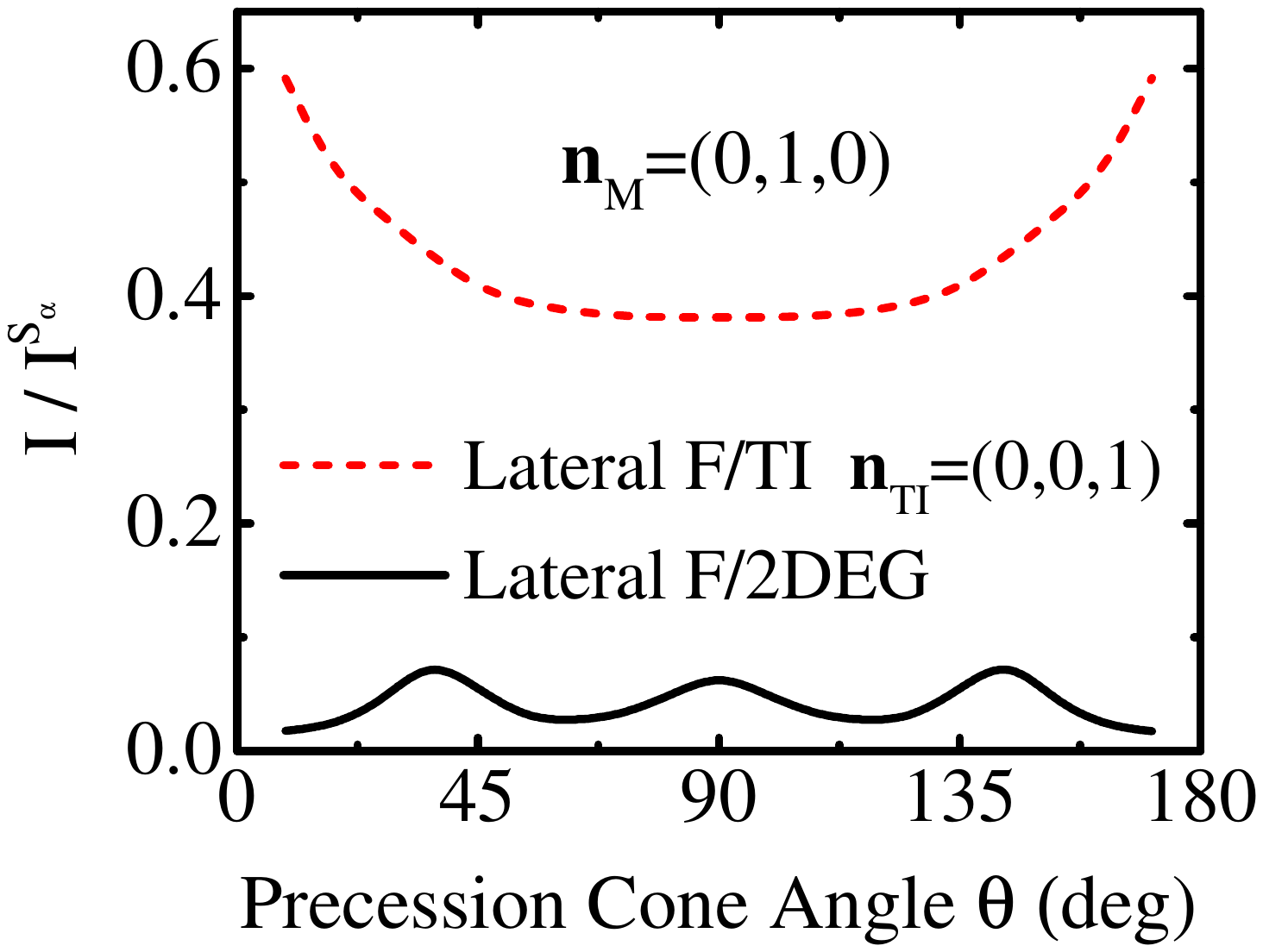}
\caption{(Color online) Efficiency of spin-to-charge conversion in lateral F/TI and F/2DEG heterostructures is quantified by the ratio of charge
$I$ and spin $I^{S_\alpha}$ currents from Fig.~\ref{fig:fig2} for magnetization precessing around the $y$-axis [$\mathbf{n}_M=(0,1,0)$] in Fig.~\ref{fig:fig1}(a).}
\label{fig:fig3}
\end{figure}

These observations have been interpreted as the ``inverse Edelstein effect''~\cite{Shen2014} (IEE) where {\em nonequilibrium} spin density in diffusive two-dimensional electron gas (2DEG) which is spin-split by SOC creates an electric field perpendicular to the spin direction. Such electric field then drives charge current $I$, or induces the dc voltage signal $V_\mathrm{pump}(\theta)=I(\theta)/G(\theta)$ in an open circuit where $G(\theta)$ is the conductance of a two-terminal system with static magnetization. In the case of heterostructures in Fig.~\ref{fig:fig1}(a), the required nonequilibrium spin accumulation arises due to spin current $I^{S_\alpha}$ pumped along the $z$-axis by the precessing F layer, which is {\em diverted} to flow within the interface because of highly resistive Bi or TI layers. Such spin current along the $x$-axis has a dc component carrying spins pointing in the $y$-direction, so that charge current observed also along the $x$-axis is unrelated to the inverse SHE. It has been speculated~\cite{Shiomi2013} that the efficiency of conversion in the case of F/TI heterostructures could reach its maximum $I/I^{S_\alpha} \rightarrow \infty$ due to spin-momentum locking along the single circle~\cite{Hasan2010} (formed in $k$-space at the intersection of the Dirac cone energy-momentum dispersion and the Fermi energy plane), in contrast to spin-momentum locking along than two circles~\cite{Sanchez2013,Shen2014,Winkler2003} in the case of conventional massive electrons described by Rashba Hamiltonian in Eq.~\eqref{eq:rashba} which counter the effect of each other.

However, these explanations operate~\cite{Shiomi2013,Shen2014} with nonequilibrium spin density, rather than with time-dependent pumped
spin current which is giving rise to it, so it remains unclear how much of it is actually converted into charge current and how efficient are
different types of SOC in this conversion process.~\cite{Takeuchi2010,Ohe2008} Another mechanism of pumped-spin-to-charge conversion was predicted in Ref.~\onlinecite{Mahfouzi2012} for {\em vertical} F/I heterostructures, such as those illustrated in Fig.~\ref{fig:fig1}(b) using F/TI system, where pumped spins are reflected and transmitted {\em perpendicularly} through the interface with strong SOC which leads to charge current (or voltage $V_\mathrm{pump}$ in open circuit) along the $x$-axis in Fig.~\ref{fig:fig1}(b).

Here we provide a unified treatment for both of these pumped-spin-to-charge conversion effects using Floquet-nonequilibrium Green
function (Floquet-NEGF) formalism~\cite{Mahfouzi2012,Chen2013} applied to time-dependent Hamiltonian of lateral and vertical heterostructures depicted in Figs.~\ref{fig:fig1}(a) and ~\ref{fig:fig1}(b), respectively, assuming ballistic transport regime for simplify. For lateral heterostructures in Fig.~\ref{fig:fig1}(a), we demonstrate in Fig.~\ref{fig:fig2} that {\em both} charge $I$ and spin $I^{S_\alpha}$ currents will flow  within the plane of F/TI or F/2DEG interface in the $x$-direction. The charge current in Figs.~\ref{fig:fig2}(a) and ~\ref{fig:fig2}(c) is non-zero only when magnetization is precessing around the $y$-axis---this setup injects dc component of spin current into the interface with spins pointing along the $y$-axis, which is partially converted into charge current along the $x$-axis. On the other hand, when F layer magnetization is precessing around the $x$- or the $y$-axis, the charge current along the $x$-axis is identically zero $I \equiv 0$, while non-zero pure spin currents $I^{S_x}$ and $I^{S_y}$ continue to
flow along the $x$-axis, as shown in Figs.~\ref{fig:fig2}(b) and ~\ref{fig:fig2}(d).

While this picture is fully compatible with the one based on IEE,~\cite{Shen2014} our approach finding both charge and spin currents makes it possible to quantify the spin-to-charge conversion efficiency for which we employ the ratio $I/I^{S_\alpha}$.  To quantify total spin angular momentum emitted by the central region of systems in Fig.~\ref{fig:fig1} into two normal metal (N) leads, we sum up the magnitudes of spin currents in the left (L) and the right (R) lead to get $I^{S_\alpha}=|I^{S_\alpha}_L|+|I^{S_\alpha}_R|$. Note that we employ the same units for charge current, $I=I^\uparrow + I^\downarrow$, and spin current, $I^{S_\alpha} = I^\uparrow - I^\downarrow$, expressed in terms of spin-resolved charge currents $I^\uparrow$ and $I^\downarrow$ carrying spins pointing along the $\alpha$-axis. Thus, the ratio $I/I^{S_\alpha}$ is a pure number which is plotted in Fig.~\ref{fig:fig3}. It reaches $I/I^{S_\alpha} \simeq$ 2--8\% for F/2DEG interface, and increases to  $I/I^{S_\alpha} \simeq$ 40--60\% for F/TI interface. We note that in the case of conventional spin-polarized charge current, $0< I^{S_\alpha}/I \le 1$ is component of the spin-polarization vector along the $\alpha$-axis. In Fig.~\ref{fig:fig3} this number can be bigger than one because spin current was initially generated by pumping in the absence of dc bias voltage and without any net charge flux, and subsequently only partially converted into charge current by the SO-coupled interface.

\begin{figure}
\includegraphics[scale=0.34,angle=0]{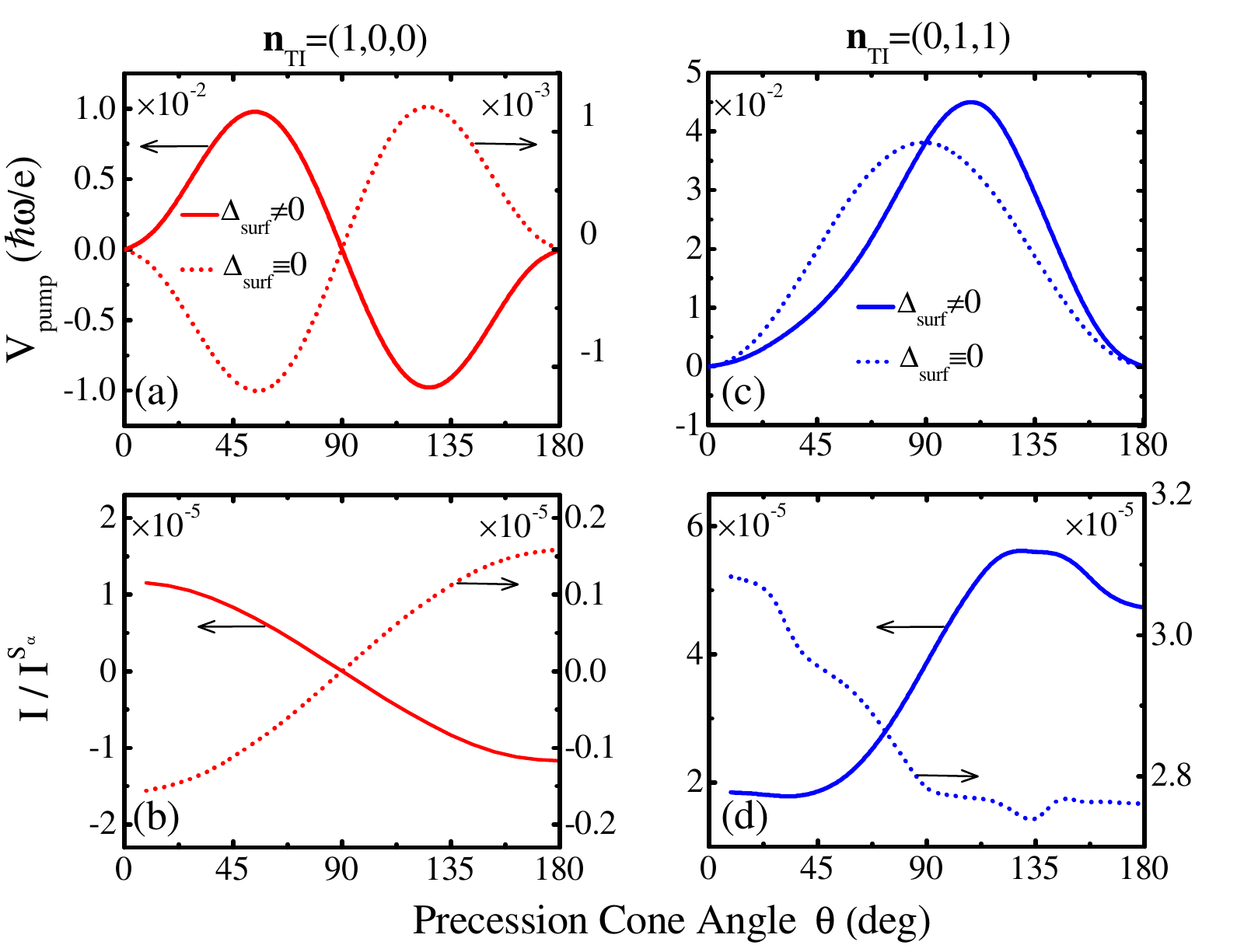}
\caption{(Color online) (a), (c) The angular dependence of the dc pumping voltage in F/TI vertical heterostructures illustrated in Fig.~\ref{fig:fig1}(b) whose magnetization is precessing around the $z$-axis, $\mathbf{n}_M=(0,0,1)$. Panels (b) and (d) show the spin current $I^{S_\alpha}$ which accompanies the charge current $I = V_\mathrm{pump} G$ in (a) and (c), respectively, where both $I^{S_\alpha}$ and $I$ flow along the $x$-axis in Fig.~\ref{fig:fig1}(b). The QLs of 3D TI slab are oriented perpendicular to $\mathbf{n}_\mathrm{TI}=(1,0,0)$ in panels (a) and (b), or perpendicular to $\mathbf{n}_\mathrm{TI}=(0,1,1)$ in panels (c) and (d).}
\label{fig:fig4}
\end{figure}
\begin{figure}
\includegraphics[scale=0.32,angle=0]{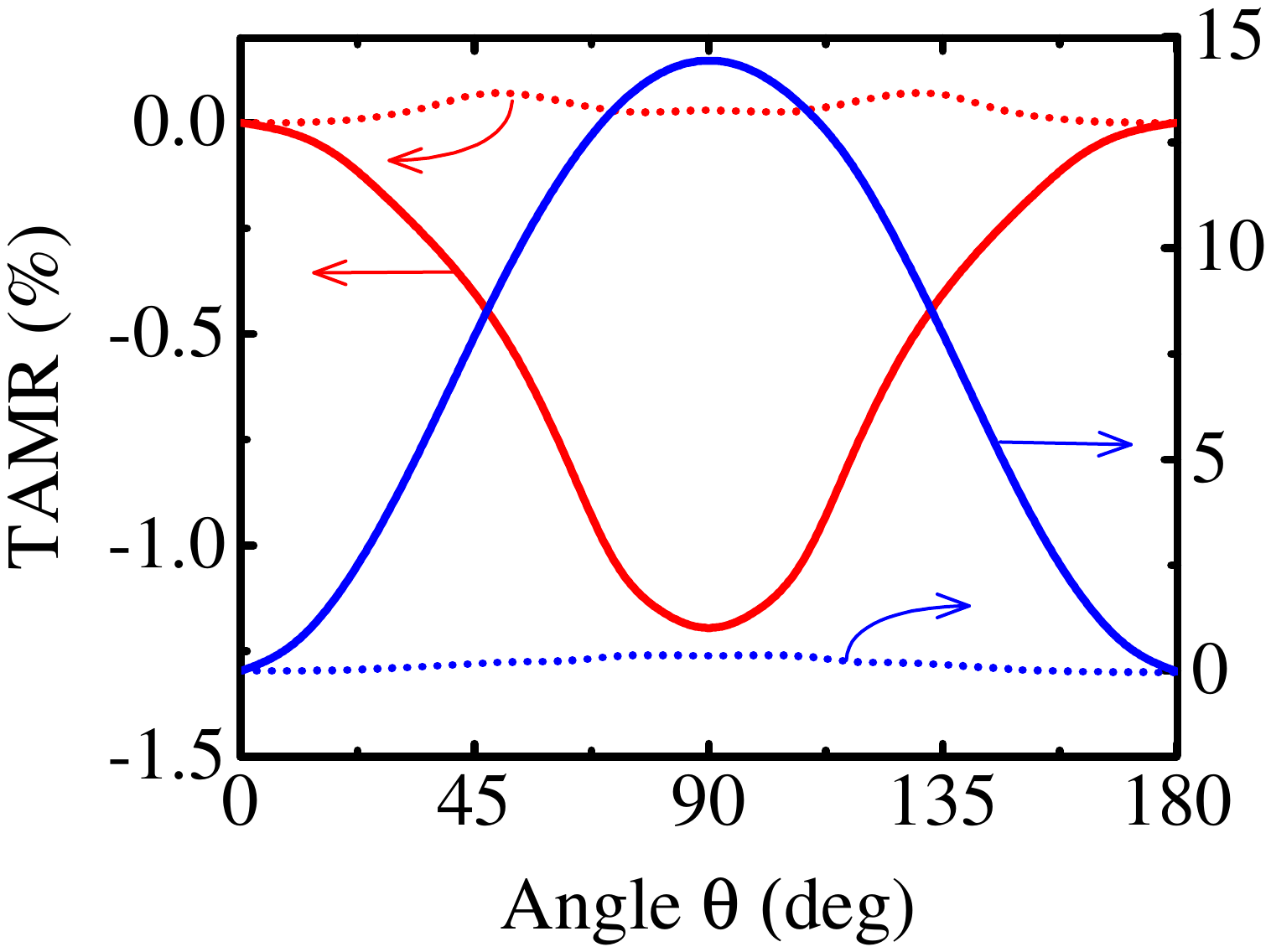}
\caption{(Color online) The TAMR of F/TI vertical heterostructures illustrated in Fig.~\ref{fig:fig1}(b) for gapless ($\Delta_{\rm surf}\equiv 0$ for dotted lines) or  gapped ($\Delta_{\rm surf} \neq 0$ for solid lines) surface of TI that is in direct contact with the F layer. The QLs of 3D TI slab are oriented perpendicular to $\mathbf{n}_\mathrm{TI}=(1,0,0)$ for curves with the left ordinate, or perpendicular to $\mathbf{n}_\mathrm{TI}=(0,1,1)$ for curves with the right ordinate.}
\label{fig:fig5}
\end{figure}

Figures~\ref{fig:fig4}(a) and ~\ref{fig:fig4}(c) shows charge current along the $x$-axis in vertical heterostructures depicted in Fig.~\ref{fig:fig1}(b) whose magnetization is precessing along the $z$-axis. In conventional F/N multilayers, magnetization dynamics pumps time-dependent pure spin current  into the N layer in the absence of any bias voltage which has been amply explored~\cite{Tserkovnyak2005} as a robust and ubiquitous example of adiabatic pumping effect at room temperature. However, no charge pumping at the adiabatic level, i.e., with contribution proportional to $\omega$, is expected in multilayers with a {\em single} precessing F layer.~\cite{Chen2009} In fact, in the absence of interfacial SOC F/I junctions can pump~\cite{Chen2009} charge current $\propto \omega^2$  which is, therefore, nonadiabatic and much smaller effect. This outcome changes if strong SOC is present directly at the interface, as predicted to occur in vertical F/I junctions with the Rashba SOC at the interface.~\cite{Mahfouzi2012} While the angular dependence of pumped charge current for F/TI interface in Fig.~\ref{fig:fig4}(a) is the same $\propto \sin^2 \theta \cos \theta$ as for the F/I interface with the Rashba SOC, the magnitude of the voltage signal is very sensitive to opening of time-dependent energy gap $\Delta_{\rm surf}(t)$ on the surface of TI by the magnetic proximity effect. That is, as soon as the cone angle $\theta$ becomes non-zero due to microwave absorption, the time-dependent exchange field acquires a component $(\Delta \sin \theta \sin \omega t){\bf e}_x$ which is perpendicular to the surface of the TI and induces the corresponding surface gap $\Delta_{\rm surf}(t)$. The value of $\Delta_{\rm surf}^{\rm max}$ is not necessarily related to $\Delta |\sin \theta|$ because magnetic proximity effect is influenced by the microscopic details~\cite{Luo2013} at the F/TI interface, so that in Fig.~\ref{fig:fig4} we consider both $\Delta_{\rm surf}(t) \neq 0$ and $\Delta_{\rm surf}(t) = 0$ cases.

The recent theoretical~\cite{Luo2013} and experimental~\cite{Yang2013} efforts have vigorously pursued F/TI heterostructures with non-zero (time-independent)  $\Delta_{\rm surf}$, as well as without complicated hybridization~\cite{Luo2013} of bulk and surface states so that split Dirac-cone remains easily identifiable. Such time-reversal-breaking-induced gapped surface state of 3D TI hosts massive Dirac fermions which make possible experimental probing of generic properties of 3D TIs like topological magnetoelectric effect (where magnetization is generated by an electric field with a quantized coefficient), half-integer quantum Hall effect and magnetic monopole.~\cite{Hasan2010} It has also been predicted~\cite{Ueda2012} that F/TI heterostructure in Fig.~\ref{fig:fig1}(a) with precession axis lying within the TI surface will pump charge current which jumps abruptly at times when the $z$-component of the precession magnetization touches zero, which is closely related to the parity anomaly in high energy physics. However, observation of such effects requires perfectly insulating bulk of TI and Fermi energy tuned close to the Dirac point (DP). On the other hand, sensitivity of charge current in Fig.~\ref{fig:fig4}(a) on the presence of $\Delta_{\rm surf}(t) \neq 0$ does not require either of these two conditions.

The Bi$_2$Se$_3$ realization of TI is a strongly anisotropic material composed of  quintuple layers (QLs) of Bi and Se atoms, where one QL consists of three Se layers strongly bonded to two Bi layers in between.~\cite{Hasan2010} The electrons on the metallic surface of Bi$_2$Se$_3$ are often described by the effective  Hamiltonian in Eq.~\eqref{eq:dirac} which accounts for spin-orthogonal-to-momentum locking for both Bi and Si sublattices observed in spin-ARPES experiments.~\cite{Hasan2010} However, such description is valid only when the surface of the TI crystal coincides with the plane of QL, while for other orientations of QLs the two sublattices generate different spin textures.~\cite{Silvestrov2012,Chang2014a} To illustrate their effect, Fig.~\ref{fig:fig4}(c) plots charge current when QLs are oriented perpendicularly to the vector $\mathbf{n}_\mathrm{TI} = (0,1,1)$ drawn in Fig.~\ref{fig:fig1}. The corresponding ratios $I/I^{S_\alpha}$ are plotted in Figs.~\ref{fig:fig4}(b) and ~\ref{fig:fig4}(d) for two different orientations of QLs (denoted on the top of the left and the right column in Fig.~\ref{fig:fig4}).

The non-zero pumping voltage $\propto \omega$ in vertical F/I junctions with interfacial SOC~\cite{Mahfouzi2012}  is closely related to the tunneling anisotropic magnetoresistance (TAMR). The out-of-plane TAMR for F/TI vertical heterostructures is defined~\cite{Mahfouzi2012} as $\mathrm{TAMR}\,(\theta) = [R(\theta) - R(0^\circ)]/R(0^\circ)$ using conventional dc resistances $R(\theta)$ measured by tilting the static magnetization of the F layer in Fig.~\ref{fig:fig1}(b) away from the $z$-axis. The TAMR curves in Fig.~\ref{fig:fig5} show that change in the conductance $G(\theta)=1/R(\theta)$ is too small to account for the enormous difference between $V_{\rm pump}(\Delta_{\rm surf} \neq 0)$ and $V_{\rm pump}(\Delta_{\rm surf} \equiv 0)$ cases in Fig.~\ref{fig:fig4}(a). Although TAMR also differentiates between the presence or absence of massive Dirac fermions on the surface of the TI,  it cannot be used reliably to detect them since it would diminish when bulk charge carriers are present within the TI slab.

\begin{figure}
\includegraphics[scale=0.3,angle=0]{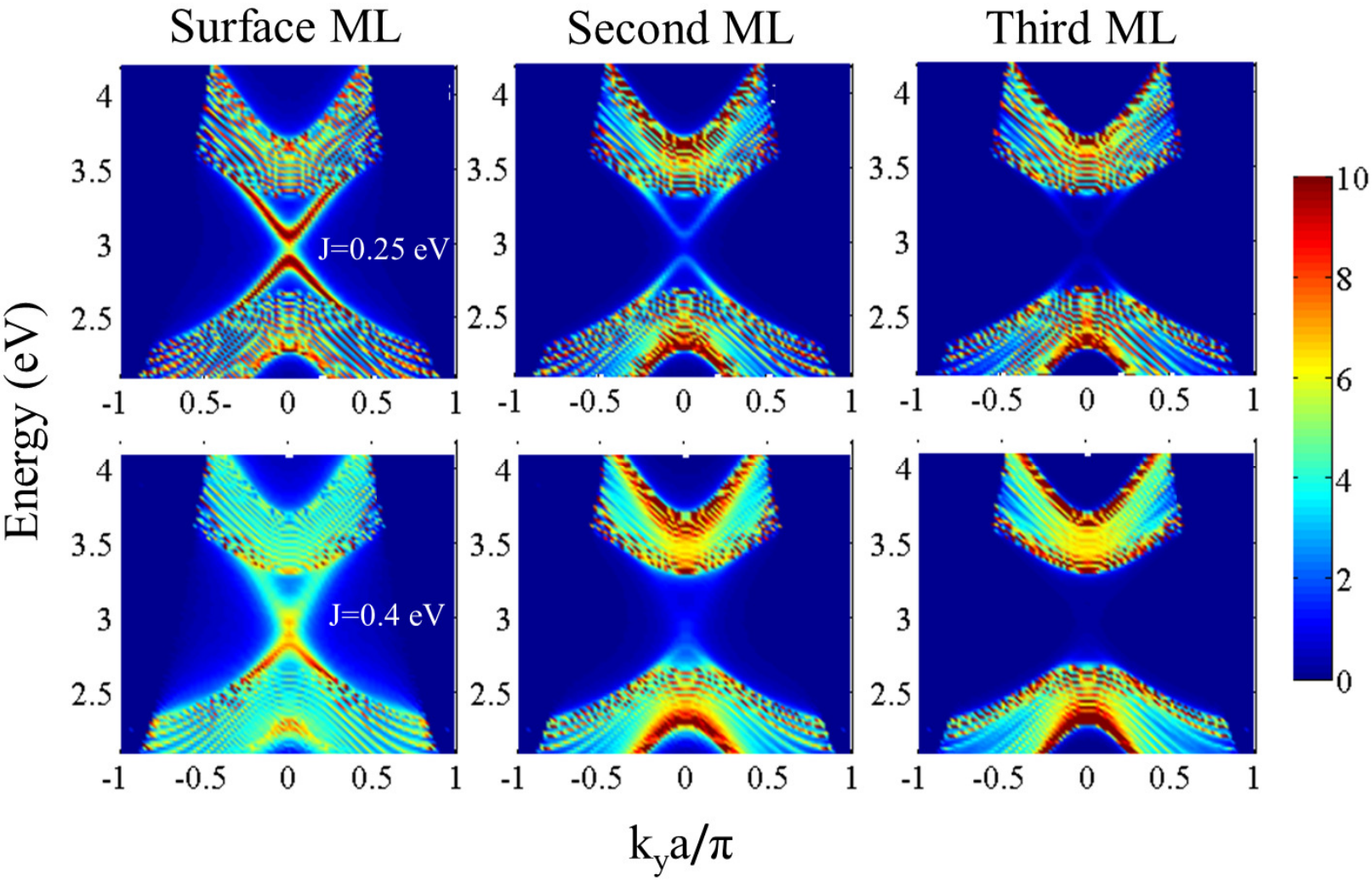}
\caption{(Color online) The local DOS $g(E,k_y,k_z=0)$ on the surface ML of TI slab in contact with the F layer within the vertical heterostructure illustrated in Fig.~\ref{fig:fig1}(b), with QLs of TI slab oriented perpendicular to $\mathbf{n}_\mathrm{TI}=(1,0,0)$, as well as on the adjacent second and third MLs. In the top row of panels, the TI slab is weakly (F-TI hopping $J=0.25$ eV) coupled to F layer, and in the bottom row of panels the coupling (F-TI hopping $J=0.4$ eV) is stronger. The static magnetization of the F layer is perpendicular to the F/TI interface, and it is assumed to induce an energy gap on the surface ML of TI via the magnetic proximity effect.}
\label{fig:fig6}
\end{figure}

We now explain details of our approach to calculating pumped spin and charge currents. The F/TI and F/2DEG lateral heterostructures illustrated in Fig.~\ref{fig:fig1}(a) are modeled on a simple cubic or square tight-binding lattice with lattice spacing $a$, respectively. The TI central region has finite length $L_x^\mathrm{TI}=50a$ and thickness $L_z^\mathrm{TI}=8a$, while it is assumed to be infinite in the $y$-direction, with each site hosting four spin-dependent orbitals of the minimal effective Hamiltonian.~\cite{Liu2010} The 2DEG central region has finite length $L_x^\mathrm{2DEG}=100a$, while it is assumed to be infinite in the $y$-direction, with each site hosting two spin-dependent orbitals of the discretized version~\cite{Mahfouzi2013} of Hamiltonian in Eq.~\eqref{eq:rashba}. We do not model explicitly the presence of the F overlayer with precessing magnetization in Fig.~\ref{fig:fig1}(a), but instead add term $-\Delta \mathbf{m}(t) \cdot {\bm \sigma}/2$ in the region of length  $L^\mathrm{F}_x$ residing in the center of the top plane of TI or plane of 2DEG.
Here $\Delta$ is the mean-field exchange splitting induced by the F overlayer through magnetic proximity effect and $\mathbf{m}(t)$ is the unit vector along the
direction of precessing magnetization. For the case of F/TI heterostructure, we select $L^\mathrm{F}_x=20a$ and $\Delta=0.28$ eV, while in the case of F/2DEG heterostructures we select $L^\mathrm{F}_x=50a$, $\Delta=0.2$ eV and $\alpha_\mathrm{R}/2a=0.1$ eV.

The vertical F/TI heterostructures in Fig.~\ref{fig:fig1}(b) are modeled on the simple cubic lattice composed of 2D monolayers (MLs) that are infinite in the $yz$-plane. The thickness of the F layer is $L_x^\mathrm{F}=50$ MLs and of the TI layer it is $L_x^\mathrm{TI}=5$ MLs. In the case of vertical heterostructures, we add  $-\Delta_\mathrm{surf} \mathbf{m}(t) \cdot {\bm \sigma}/2$ on the first monolayer of TI layer in contact with F layer, where both $\Delta_\mathrm{surf} \equiv 0$ and $\Delta_\mathrm{surf} \equiv \Delta$ ($\Delta=0.28$ eV is the exchange splitting in the bulk of the F layer) are considered in Figs.~\ref{fig:fig4} and ~\ref{fig:fig5}.

Figure~\ref{fig:fig6} shows the local density of states (LDOS) on the MLs of TI that are the closest to F layer within the vertical heterostructure in Fig.~\ref{fig:fig1}(b). When the hopping parameter between the lattice sites of F and TI surface MLs is large ($J>0.5$ eV), the energy-momentum dispersion displayed in the bottom row of Fig.~\ref{fig:fig6} bears little resemblance to the Dirac cone because of the flooding~\cite{Zhao2010} of F/TI interface by evanescent wavefunctions which originate from the F layer and penetrate into the bulk gap of TI while exponentially decaying in space. To evade this effect, we assume smaller hopping \mbox{$J=0.25$ eV} which leads to LDOS shown in the top row of panels in Fig.~\ref{fig:fig6}.  In addition, Fig.~\ref{fig:fig6} demonstrates how gapped surface state of TI can penetrate into the bulk of TI as evanescent wavefunction decaying over the first few MLs, which effectively dopes the bulk.

The explicit expressions for the Hamiltonians discussed above, including how to properly couple Hamiltonian of semi-infinite N leads or F layer with two orbitals per site to the TI Hamiltonian with four orbitals per site, can be found in Ref.~\onlinecite{Chang2014a} for the TI central region or in Ref.~\onlinecite{Mahfouzi2013} for the 2DEG central region. The bottom of the band of the TI is shifted by \mbox{$3.0$ eV}, so that DP is at $3.0$ eV and the Fermi energy is chosen at $E_F=3.1$ eV  both lateral and vertical F/TI heterostructures in Fig.~\ref{fig:fig1}. The Fermi energy for lateral F/2DEG heterostructures is set at \mbox{$E_F=3.5$ eV}.

Since pumped charge current and all components of pumped spin current tensor are time-dependent in the presence of SOC,~\cite{Mahfouzi2012,Ohe2008} the NEGF formalism~\cite{Stefanucci2013} is advantageous choice for their computation because it gives from the outset experimentally measurable current averaged over one period. Alternative scattering matrix approach to adiabatic pumping~\cite{Brouwer1998,Tserkovnyak2005} requires to compute current at all times (or, in practice, over a discrete time grid) during one period of microwave oscillations and then find its average,~\cite{Ohe2008,Hals2010} which can be very expensive computationally [especially for tunneling structures, like vertical F/TI ones in Fig.~\ref{fig:fig1}(b), where current amplitude is several orders of magnitude larger than its average value].

We compute pumped spin and charge currents in heterostructures in Fig.~\ref{fig:fig1} using the recently derived~\cite{Mahfouzi2012} {\em exact} multiphoton solution to double-Fourier-transformed NEGFs in the presence of external time periodic potential, which is adapted below to systems that are infinite in one or two directions. In terms of the operators $\hat{c}_{n \sigma}^\dag$ ($\hat{c}_{n \sigma}$) which create (annihilate) electron with spin $\sigma$ on lattice site $n$, the two fundamental objects~\cite{Stefanucci2013} of the NEGF formalism are the retarded  $G^{r,\sigma\sigma'}_{nn'}(t,t')=-i \Theta(t-t') \langle \{\hat{c}_{n\sigma}(t) , \hat{c}^\dagger_{n'\sigma'}(t')\}\rangle$ and the lesser $G^{<,\sigma\sigma'}_{nn'}(t,t')=i \langle \hat{c}^\dagger_{n'\sigma'}(t') \hat{c}_{n \sigma}(t)\rangle$ GF which depend on two time arguments. The former describe the density of available quantum states, while the latter describes how electrons occupy those states. Here $\langle \ldots \rangle$ denotes the
nonequilibrium statistical average.~\cite{Stefanucci2013}

When the system Hamiltonian depends on time explicitly one has to work with both times. To solve the equation of motion for $\hat{G}^{r,\sigma\sigma'}_{nn'}(t,t')$ and the Keldysh integral equation for  $G^{<,\sigma\sigma'}_{nn'}(t,t')$ it is advantageous to switch to a more convenient representation ($\hbar =1$)
\begin{equation}\label{eq:doubleft}
{\bf G}^{r,(<)}(t,t')=\int\limits_{-\infty}^{+\infty} \frac{dE}{2\pi} \! \int\limits_{-\infty}^{+\infty}  \frac{E'}{2\pi} \, e^{-i E t + i E' t'} {\bf G}^{r,(<)}(E,E').
\end{equation}
Due to the Floquet theorem, the double-time Fourier transformed retarded GF must take the form \mbox{${\bf G}^r(E,E')={\bf G}^r(E,E+N\omega)={\bf G}^r_N(E)$}. The coupling of energies $E$ and $E+ N\omega$, where $N$ is integer, signifies how multiphoton processes assist the generation of pumped current. Since the heterostructures considered in Fig.~\ref{fig:fig1} are translationally invariant in the transverse directions, we additionally perform spatial Fourier transform, so that multiphoton retarded GF depends on ${\bf k}_\parallel=(k_y,k_z)$ or ${\bf k}_\parallel = k_y$ in the case of heterostructures in Fig.~\ref{fig:fig1}(a) and Fig.~\ref{fig:fig1}(b), respectively.

In the absence of inelastic processes, solving $[E \check{\textbf{1}} + \check{\bm \Omega} - \check{\textbf{H}}_{{\bf k}_\parallel} - \check{\bm \Sigma}^r_{{\bf k}_\parallel} (E + \check{\bm \Omega})] \check{\textbf{G}}^r_{{\bf k}_\parallel} (E) = \check{\textbf{1}}$ for the multiphoton retarded GF is sufficient to obtain  the time-averaged pumped spin current in N lead $p=L,R$ as:
\begin{eqnarray}\label{eq:central}
 I^{S_\alpha}_p & = &  \frac{e}{2 N_{\rm ph} } \int d{\bf k}_\parallel \, {\rm Tr}\, \{ \hat{\sigma}_{\alpha} \check{\bf{\Gamma}}_{p,{\bf k}_\parallel}  [\check{\bf{\Omega}},\check{\textbf{G}}^r_{{\bf k}_\parallel}(E_F) \check{\bf{\Gamma}}_{{\bf k}_\parallel}(E_F)] \}, \nonumber \\
 && [\check{\bf{\Omega}},\check{\textbf{G}}^r_{{\bf k}_\parallel} (E_F) \check{\bf{\Gamma}}_{{\bf k}_\parallel}(E_F)]  =  \check{\bf{\Omega}}\check{\textbf{G}}^r_{{\bf k}_\parallel}(E_F)\check{\bf{\Gamma}}_{{\bf k}_\parallel}(E_F) \nonumber \\
 && - \check{\textbf{G}}^r_{{\bf k}_\parallel}(E_F) \check{\bf{\Gamma}}_{{\bf k}_\parallel}(E_F) \check{\bf{\Omega}},
\end{eqnarray}
where the adiabatic limit $\omega \ll E_F$ satisfied by the frequency of microwaves is taken into account. By replacing the Pauli matrix $\hat{\sigma}_{\alpha}$ in Eq.~\eqref{eq:central} with the unit $2 \times 2$ matrix, one can also obtain the pumped charge current $I_p$. We use the symbol $\check{\bf A}$ to denote matrices which act in the total Hilbert space ${\mathcal H}_{\rm el} \otimes {\mathcal H}_{\rm ph}$, where the dimension of the Hilbert space of photons ${\mathcal H}_{\rm ph}$ is infinite and the dimension of ${\mathcal H}_{\rm el}$ is equal to the number of atomic orbitals comprising the central region. The unit matrix in the Hilbert space of a single electron ${\mathcal H}_{\rm el}$ is ${\bf 1}$, and the unit matrix in ${\mathcal H}_{\rm el} \otimes {\mathcal H}_{\rm ph}$ is denoted by $\check{\mathbf 1}$. In Eq.~\eqref{eq:central}, $\check{\bm \Omega}={\rm diag} \, (\cdots, -2\omega{\bf 1}, -\omega{\bf 1}, 0, \omega{\bf 1}, -2\omega{\bf 1}, \cdots)$ is diagonal matrix and $\check{\bm \Gamma}_{{\bf k}_\parallel} = \sum_p \check{\bm \Gamma}_{p,{{\bf k}_\parallel}}$ is the level broadening matrix due to the coupling of the central region in Fig.~\ref{fig:fig1} to N leads. Because the trace in the integrand, ${\rm Tr} \equiv {\rm Tr}_{\rm el} {\rm Tr}_{\rm ph}$, is summing over contributions from different photon exchange processes, the denominator includes $2N_{\rm ph}$  to avoid double counting. Importantly, the part of the trace operating in ${\mathcal H}_{\rm ph}$ space ensures the charge current conservation in our solution~\cite{Mahfouzi2012} to NEGF equations. We find that finite number  $N_{\rm ph}=10$ of exchanged microwave photons is sufficient to obtain converged values for pumped spin and charge current.

In conclusion, inspired by the recent experiments~\cite{Sanchez2013,Shiomi2013} observing conversion of pure spin current pumped by microwave-driven precessing magnetization into charge current flowing within the F/2DEG or F/TI interface---where conversion is due to strong interfacial SOC but it is unrelated to widely  employed~\cite{Saitoh2006,Mosendz2010,Ando2011a} ISHE for detection of pure spin currents---we provide a unified theoretical treatment of this process together with spin-to-charge conversion when pumped spins flow perpendicularly through the SO-coupled interfaces. Our approach is based on charge-conserving Floquet-Green function formalism~\cite{Mahfouzi2012,Chen2013} for open quantum systems driven by external time periodic potential, which makes possible to quantify conversion efficiency $I/I^{S_\alpha}$ by computing {\em both} the charge and spin currents outflowing from such systems into the macroscopic
reservoirs (such information has eluded recent alternative explanation of the experiments in Refs.~\onlinecite{Sanchez2013,Shiomi2013} based on the nonequilibrium spin density-driven IEE~\cite{Shiomi2013,Shen2014}). The highest conversion efficiency $I/I^{S_\alpha} \simeq$ 40--60\% (depending on the precession cone angle) occurs when spins are pumped into the F/TI interface due to perfect spin-momentum locking on the surface of 3D TIs. On the other hand, even though conversion efficiency is orders of magnitude smaller for pumping vertically through the F/TI interface, we predict that converted charge current could be used as a sensitive probe for the presence of massive Dirac fermions at this interface which does not require perfectly insulating bulk of TI or Fermi energy tuned into the surface energy gap.

\begin{acknowledgments}
F. M. and B. K. N. were supported by NSF under Grant No. ECCS 1202069. N. N. was supported by Grant-in-Aids for Scientific Research (No. 24224009) from the Ministry of Education, Culture, Sports, Science and Technology of Japan, Strategic International Cooperative Program (Joint Research Type) from Japan Science and Technology Agency, and also by Funding Program for World-Leading Innovative
R\&D on Science and Technology (FIRST Program).
\end{acknowledgments}


\end{document}